\documentclass{PoS}

\title{Comments on staggered fermions / Panel discussion}

\ShortTitle{Comments on staggered fermions / Panel discussion}

\author{\speaker{Michael Creutz}%
         \thanks{This manuscript has been
authored under contract number DE-AC02-98CH10886 with the
U.S.~Department of Energy.  Accordingly, the U.S. Government retains a
non-exclusive, royalty-free license to publish or reproduce the
published form of this contribution, or allow others to do so, for
U.S.~Government purposes.}\\
        Brookhaven National Laboratory\\
        E-mail: \email{creutz@bnl.gov}}

\abstract{The rooting procedure commonly
used with staggered fermions does not correctly treat non-perturbative
effects associated with gauge field topology.  In practice these
effects are small for the physics of flavor non-singlet particles.
However large uncontrolled systematic errors are expected for flavor
singlet issues, such as the mass of the eta prime meson.  While the
relative speed of the algorithm in large scale simulations may justify
its use, the method is an approximation and should not be promoted as
a first principles approach to the strong interactions.  }

\FullConference{8th Conference Quark Confinement and the Hadron Spectrum\\
        September 1-6, 2008\\
        Mainz. Germany}

\begin{document}


Within the lattice gauge community there currently exists a rather
bitter ongoing controversy over whether one should be using the rooted
staggered fermion algorithm for large scale simulations of QCD.  While
the approach has been shown to be a good approximation for many
quantities, it mistreats anomaly effects and thus is not a first
principles approach to the strong interactions.  Here I summarize the
issues for the broader audience at this conference.

As all here are familiar, chiral symmetry plays a key role in our
understanding of the strong interactions.  From a conceptual point of
view, pions are most elegantly described as waves propagating on a
quark condensate, $\langle\overline\psi\psi\rangle$.  And from a
practical point of view, chiral extrapolations are an essential tool
for the lattice gauge theorist to extract predictions from simulations
with quark masses heavier than their physical values.

The subject of chiral symmetry is intricately entwined with quantum
anomalies.  As is well known, these remove the classical $U(1)$ chiral
symmetry of the theory.  For $N_f$ massless quark flavors, the
surviving chiral symmetry is $SU(N_f)_L\times SU(N_f)_R$.  Since
$SU(1)$ is trivial, a useful chiral symmetry requires at least two
flavors.  Indeed, an exact chiral symmetry for two flavors is possible
on the lattice using minimally doubled actions \cite{minimally}, but
that is not the subject of this session.

Ignoring the anomaly in lattice gauge theory frequently leads to what
are known as doublers.  For example, the most naive fermion action has
quarks hopping from site to site picking up gamma matrix factors.
That approach maintains chiral symmetry, but on further analysis
actually describes sixteen fermions in the continuum limit.  The
formulation possesses an exact $U(4)\times U(4)$ chiral symmetry
\cite{Karsten:1980wd}.  Dividing out that symmetry gives rise to the
staggered fermion approach.  This leaves behind four remaining
species, often called tastes, while maintaining one exact chiral
symmetry \cite{Kogut:1974ag,Susskind:1976jm,Sharatchandra:1981si}.
The alternative Wilson fermion approach does succeed in eliminating
all doubling, but at the expense of breaking any remnants of chiral
symmetry at finite lattice spacing \cite{Wilson:1975id}.  Other
techniques, such as perfect actions \cite{perfect}, domain wall quarks
\cite{Kaplan:1992bt}, or the overlap operator \cite{Neuberger:1997fp},
do maintain a form of chiral symmetry for any $N_f$, although these
all involve interactions over a range of sites and thus are extremely
computationally intensive.  Furthermore, with these actions the
anomaly is often somewhat hidden; for example, the overlap formalism
introduces two different chiral matrices $\gamma_5$ and $\hat
\gamma_5$, with the latter being non-local, gauge field dependent, and
having a trace related to the winding number of the background gauge
field.  The minimally doubled actions do maintain a strict locality
along with one exact chiral symmetry, but suffer from potential
lattice distortions that should be tuned at finite lattice spacing.

The issue at the heart of the recent controversies over staggered
fermions is the process referred to as rooting.  In an attempt to
remove the fourfold multiplicity of the staggered approach,
simulations are done with the determinant of the staggered operator
replaced with its fourth root.  In a perturbative expansion this
multiplies each fermion loop by 1/4, correcting for the extra
multiplicity of the four tastes.  Thus rooting is a valid procedure in
perturbation theory.  But all symmetries of the fermion determinant
are kept by its fourth root, and the one flavor theory has no chiral
symmetry.  Thus something must go wrong at a non-perturbative level.

At last year's lattice meeting in Regensburg I went into some detail
as to how this rooting procedure fails.  Those that are interested in
the specifics can refer to Ref.~\cite{Creutz:2007rk}.  Further
discussion can be found in Refs.~\cite{Creutz:2007yg} and
\cite{Creutz:2007yr}. Here I discuss the qualitative essence of the
problem, which is that the rooting procedure does not treat instanton
effects properly.

\begin{figure*}
\centering
  \includegraphics[height=.15\textheight]{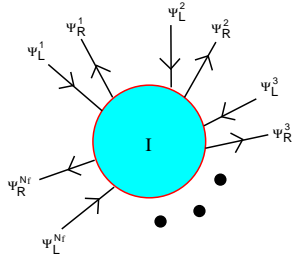}
\caption{A topologically non-trivial gauge field generates an
effective interaction between all fermion species, each of which flips
its spin.}
\label{instanton}
\end{figure*}

Whenever the gauge fields are topologically non-trivial, the continuum
Dirac operator develops real eigenvectors, and these appear for every
flavor and taste.  At zero mass the corresponding eigenvalues go to
zero.  Depending on the action, lattice artifacts can mix these modes
and make them not exactly real, but near the continuum limit they
should exist as approximately real eigenvalues.  Furthermore, in the
continuum these modes become chiral, i.e. the eigenvectors are also
eigenstates of $\gamma_5$, with the eigenvalue depending on the
winding number of the gauge field.  The physical effect of these modes
is an effective coupling of all species whenever an instanton is
present \cite{thooft}.  This is sketched in
Fig.~(\ref{instanton}). Whether or not the gauge configurations are
obtained with rooting, this coupling involves every taste at once.
The net effect eliminates the possibility of considering the tastes as
independent, and the factorization crucial to the perturbative
argument fails.  In particular, the resulting coupling between tastes
should not be present in the target theory with a reduced number of
species.

Unfortunately numerous misleading statements continue to be propagated
about this problem.  To begin with, it has nothing to do with a
breaking of the symmetry between the tastes.  The unwanted coupling
takes the form of a determinant which preserves this symmetry.
Second, the issue is not associated with taking the fermion mass to
zero.  The effective coupling is present at finite mass, and indeed is
enhanced with the mass since instantons are suppressed in the chiral
limit.  Finally, and most important, the problematic coupling does not
go away in the continuum limit.  The issues occur at the typical
instanton scale, which is set by $\Lambda_{\rm qcd}$.  Simple symmetry
arguments forbid a reduction to the desired form of the instanton
induced interaction in the target theory.

To see this more explicitly, note that with $N_f$ physical flavors, in
addition to the usual flavored chiral symmetries, massless continuum
QCD has a $Z_{N_f}$ discrete chiral symmetry \cite{Creutz:1995wf}
under
\begin{equation}
\psi_L \rightarrow e^{2\pi i/ N_f} \psi_L.
\end{equation}
Here I have chosen a left handed chiral rotation; equivalently one
could work with $\psi_R$.  This symmetry can be seen either directly
from the 't Hooft vertex, which receives a factor from each species,
or from the conventional chiral Lagrangian approach because $e^{2\pi
i/ N}$ is an element of the flavored chiral symmetry group $SU(N)$.
In terms of the singlet composite fields $\sigma\sim\overline\psi\psi$
and $\eta^\prime\sim i\overline\psi\gamma_5 \psi$ (sum over flavors
implied), the effective potential of the theory is symmetric under
\begin{eqnarray}
& \sigma      & \rightarrow \cos(2\pi/N_f)\sigma-\sin(2\pi/N_f)
  \eta^\prime  \\
& \eta^\prime & \rightarrow \sin(2\pi/N_f)\sigma+\cos(2\pi/N_f)
  \eta^\prime .
\end{eqnarray}
Note that because the anomaly generates an $\eta^\prime$ mass, the
theory must not be symmetric under $U(1)$ rotations of the above
form with angles less than $2\pi/N_f$.

Now the staggered fermion operator has $4N_f$ effective flavors.  This
means that near the continuum limit the action possesses a $Z_{4N_f}$
symmetry.  This is true for any gauge configuration, independent of
whether it is obtained by rooting or not.  The excess symmetry forbids
a reduction to the desired $Z_{N_f}$ symmetry for the rooted theory.

\begin{figure*}
\centering
  \includegraphics[height=.14\textheight]{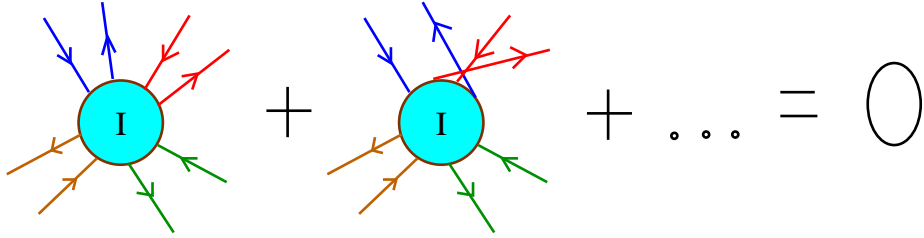}
\caption{Were one to root four copies of a single fermion, Pauli
statistics require including exchange diagrams that cancel the
analogue of the coupling between the four tastes of staggered
fermions.}
\label{pauli}
\end{figure*}

It has been argued that because of the trivial mathematical identity
$\left(|D|^4\right)^{1/4}\equiv |D|$, rooting four copies of a single
fermion should work, invalidating the above discussion.  This misses
the crucial point that staggered fermions are not four copies of one
fermion.  Indeed, as discussed in some detail in \cite{Creutz:2007rk},
the different tastes have different chiralities.  When a topological
defect is present, two tastes have a left handed zero mode and two are
right handed.  Rooting effectively averages over physically
inequivalent states.

More generally, the staggered propagator gives rise to four poles
representing the four inequivalent fermions.  Since the staggered
Dirac operator is normal, the multiple eigenvectors associated with
zero modes are exactly orthogonal.  In contrast, for the correct one
flavor theory there is only one zero mode.  With a single state, Pauli
statistics must be taken into account and will cancel any analogous
contributions from multiple copies of the single fermion, as sketched
in Fig.~(\ref{pauli}).

So rooting is wrong, but could it be a good approximation?  The answer
here appears to be yes, based on the many good past results.  This can
be partially understood from the fact that virtual quark loops are a
relatively small correction to the valence approximation.  Also, since
the problems involve all flavors and tastes at once, the issues don't
arise until an effective $N_f$ loop order.  Nevertheless, results from
rooting that involve singlet processes, where topological effects are
important, are extremely suspect.

To conclude, the basic question is ``Does expediency justify rooted
staggered quarks?''  Obviously I don't think so, but some others in
the lattice gauge community feel differently.  I summarize in Table 1
the main arguments for and against using this algorithm.
I stress, however, that if rooting is used, it is necessary to admit
that it is an approximation involving uncontrolled systematic errors
when instanton physics is important.

\vfill\eject

\begin{table}
\centering
\def \li {\hskip .2in {$\bullet$}\hskip .1 in}
\begin{tabular}{|l|l|}
\hline
\multicolumn{2}{|c|}{{\bf Pro} \hskip 2in  {\bf Con}} \\
\hline
A good approximation &  But not first principles QCD\\
\li good for flavor physics & \li incorrect singlet physics\\
 Fast simulations & Wilson fermions not far behind \cite{DelDebbio:2006cn} \\
\li thermo needs high statistics & \li twisted catching up\\
Exact chiral symmetry & Even when wrong; {\it i.e.} $N_f=1$ \\
 \li simple chiral extrapolations & \li valid chiral algorithms exist\\
Much faster than DWF & So are minimally doubled actions\\
Big lattices freely available & You get what you pay for\\
A lot already invested &  Don't throw good money after bad\\
\hline
\end{tabular}

\caption{Arguments pro and con for investigating lattice
gauge theory using rooted staggered quarks.}
\end{table}

\end{document}